\documentclass
[showpacs,amsfonts,amssymb,oneside,notitlepage,10pt,a4paper,twocolumn,prl,preprint,nobalancelastpage]{revtex4}%
\usepackage{amsfonts}
\usepackage{amsmath}
\usepackage{amssymb}
\usepackage{graphicx}%
\providecommand{\U}[1]{\protect\rule{.1in}{.1in}}
\begin{document}
\title{On canonical pairs in two-transverse-mode DOPOs}
\author{Carlos Navarrete-Benlloch, Eugenio Rold\'{a}n, and Germ\'{a}n J. de Valc\'{a}rcel}
\affiliation{Departament d'\`{O}ptica, Universitat de Val\`{e}ncia, Dr. Moliner 50,
46100--Burjassot, Spain.}

\begin{abstract}
In ref. \cite{PRL} we analyzed the properties of a Degenerate Optical
Parametric Oscillator (DOPO) tuned to the first transverse mode family at the
signal frequency. Above threshold, a Hermite-Gauss mode with an arbitrary
orientation in the transverse plane is emitted, and thus the rotational
invariance of the system is broken. When quantum effects were taken into
account, it was found on the one hand, that quantum noise is able to induce a
random rotation on this classically emitted mode. On the other hand, the
analysis of a balanced homodyne detection in which the local oscillator (LO)
was orthogonal to the excited mode at any time, showed that squeezing in the
quadrature selected by the LO was found for every phase $\psi_{\mathrm{L}}$ of
this one, squeezing being perfect for $\psi_{\mathrm{L}}=\pi/2$. This last
fact revealed an apparent paradox: If all quadratures are below shot noise
level, the uncertainty principle seems to be violated. In \cite{PRL} we stated
that the explanation behind this paradox is that the quadratures of the
rotating orthogonal mode do not form a canonical pair, and the extra noise is
transferred to the diffusing orientation. Thes notes are devoted to prove this claim.

\end{abstract}

\maketitle

\textit{Statement of the problem}. In the work of Ref. \cite{PRL} we showed
the possibility of using the spontaneous rotational symmetry breaking that
occurs in the transverse plane of some optical systems to produce
non-critically squeezed light. In particular, we proved this in the case of a
type I Degenerate Optical Parametric Oscillator (DOPO) tuned to the first
transverse mode family at the signal frequency. Let us remind what was found
there. [If you are familiar with the results showed in \cite{PRL} just jump to
Eq. (\ref{Spectrum}).]

We consider a type I DOPO pumped by a Gaussian resonant beam at frequency
$2\omega_{0}$. At the subharmonic, i.e., at $\omega_{0}$, the cavity is
assumed to be tuned to the first transverse mode family, which supports two
Laguerre-Gauss modes, $L_{\pm1}\left(  \mathbf{r}\right)  $, with opposite
orbital angular momenta: $L_{\pm1}\left(  \mathbf{r}\right)  =\pi^{-1/2}%
w^{-2}re^{-r^{2}/2w^{2}}e^{\pm i\phi}$, where $r$ and $\phi$ are the polar
coordinates in the transverse plane, and $\sqrt{2}w$ is the waist radius of
the signal beam.

The signal field operator at frequency $\omega_{0}$ can then be written as%
\begin{equation}
\hat{E}_{\mathrm{s}}\left(  \mathbf{r,}t\right)  =\hat{A}_{\mathrm{s}}\left(
\mathbf{r,}t\right)  e^{-i\omega_{0}t}+\mathrm{H.c.},
\end{equation}
apart from an unimportant constant factor, where the slowly varying envelope%
\begin{equation}
\hat{A}_{\mathrm{s}}\left(  \mathbf{r,}t\right)  =\hat{a}_{+1}\left(
t\right)  L_{+1}\left(  \mathbf{r}\right)  +\hat{a}_{-1}\left(  t\right)
L_{-1}\left(  \mathbf{r}\right)  \label{SVE}%
\end{equation}
and the interaction picture boson operators satisfy the usual canonical
conmutation relations%
\begin{equation}
\left[  \hat{a}_{i}\left(  t\right)  ,\hat{a}_{j}^{\dagger}\left(  t\right)
\right]  =\delta_{ij}\text{ \ }i,j=\pm1\text{.} \label{CanCom}%
\end{equation}

In the classical limit, the modal boson operators $\left(  \hat{a}_{j},\hat
{a}_{j}^{\dagger}\right)  $ coincide with the normal variables for each mode
$\left(  \alpha_{j},\alpha_{j}^{\ast}\right)  $, and the first result we
proved in \cite{PRL} was that the long term classical emission of the DOPO
pumped above threshold for signal modes oscillation is given by%
\begin{equation}
\bar{\alpha}_{\pm1}=\rho e^{\mp i\theta} \label{StSol}%
\end{equation}
where $\rho$ is an amplitude which depends on the system parameters and whose
exact dependence is not important for thes notes, and $\theta$ is an
\textit{arbitrary} phase, i.e., the phase difference between the $\alpha
_{\pm1}$ modes is not fixed by the classical equations of the system.

When the quantum properties of the system are analyzed, we take this
classically undefined phase $\theta$ as a quantum variable in the positive
\textit{P}-representation, finding that it diffuses in an undamped way because
of quantum noise. Hence, under these circumstances the emission of the system
will be given by (we omit the $\mathbf{r}$ dependences of the modes)%
\begin{equation}
A_{\mathrm{s}}=\rho\left[  e^{-i\theta\left(  t\right)  }L_{+1}+e^{i\theta
\left(  t\right)  }L_{-1}\right]  \propto re^{-r^{2}/2w^{2}}\cos\left[
\phi-\theta\left(  t\right)  \right]  \label{BrightMode}%
\end{equation}
i.e., the system is emitting in a Hermite-Gauss TEM$_{10}$ mode rotated by an
angle $\theta\left(  t\right)  $ with respect to the $x$-axis. The random
difussion of $\theta$ means that the orientation of the classically excited
mode is totally undetermined in the long time limit.

What was exploited in Ref. \cite{PRL} is the following idea. If the angular
orientation of the TEM$_{10}$ mode in the transverse plane is completely
undetermined, it would be natural that its associated orbital angular momentum
were completely determined, i.e., squeezed. This simple idea was proved
rigurously in \cite{PRL} by introducing a homodyne detection scheme in which,
using a 50/50 beam splitter, the field exiting the DOPO is mixed with a local
oscillator (LO) proportional to the orbital angular momentum of the
classically excited mode $\left(  -i\partial_{\theta}A_{\mathrm{s}}\right)  $,
i.e.,%
\begin{equation}
A_{\mathrm{LO}}=\rho_{\mathrm{L}}e^{i\psi_{\mathrm{L}}}\left[  e^{-i\theta
\left(  t\right)  }L_{+1}-e^{i\theta\left(  t\right)  }L_{-1}\right]
\propto\sin\left[  \phi-\theta\left(  t\right)  \right]  \label{LO}%
\end{equation}
which in this case coincides with another Hermite-Gauss mode, orthogonal to
the classically emitted one at any time. Note that $\rho_{\mathrm{L}}$ is a
real amplitude and we have denoted by $\psi_{\mathrm{L}}$ the phase of the LO.
The best way to observe squeezing is to measure the spectrum of the intensity
difference between the two output ports of the beam splitter, denoted by
$V\left(  \omega\right)  $. If $V\left(  \omega_{\ast}\right)  =0$, one can
say that the field generated by the DOPO at noise frequency $\omega_{\ast}$
has no noise in the quadrature selected by the LO, i.e. it is perfectly squeezed.

When the calculation of this spectrum is carried out one obtains \cite{PRL}%
\begin{equation}
V_{\psi_{\mathrm{L}}}\left(  \omega\right)  =1-\frac{\sin^{2}\left(
\psi_{\mathrm{L}}\right)  }{1+\left(  \omega/2\gamma_{\mathrm{s}}\right)
^{2}} \label{Spectrum}%
\end{equation}
where $\gamma_{\mathrm{s}}$ is the cavity linewidth at the signal frequency.
This result confirms what we suspected with the simple reasoning given some
lines above: The phase quadrature ($\psi_{\mathrm{L}}=\pi/2$) of the angular
momentum of the classically excited mode is perfectly squeezed at the signal
frequency ($\omega=0$). Moreover for any LO phase $V_{\psi_{\mathrm{L}}%
}\left(  \omega\right)  \leq1$, i.e., any quadrature of the selected mode has
noise reduction (except $\psi_{\mathrm{L}}=0$ where the equality of this
expression holds). This is an unexpected result, as in usual DOPOs when noise
is removed from one quadrature, it is transferred to its orthogonal quadrature
in order to fulfil the uncertainty principle, as they form a canonical pair.

In \cite{PRL} we stated that this result can be understood if the two
orthogonal \textit{detected} quadratures do not form a canonical pair, and
that this would be because the squeezed mode (and hence the LO) is rotating
randomly (because of the already commented diffusion of $\theta$). Hence, the
noise supression of any rotating quadrature is not transferred to another
quadrature, but to the orientation of the squeezed mode. If so, the undefined
orientation and the squeezed quadrature have to form the canonical pair. The
rest of the notes are devoted to prove that this intuitive explanation is
actually correct.

\textit{The way up. }The proof that reinforces the previous explanation relies
on the quantum operator which is detected by the scheme we presented some
lines above, i.e., the \textit{rotating }quadrature. By projecting the field
exiting the cavity onto the LO one can find that this operator is (we have
reversed the coherent representation for the $\theta$ variable)%
\begin{equation}
\hat{X}^{\psi_{\mathrm{L}}}=\frac{i}{\sqrt{2}}\left[  e^{-i\psi_{L}}\left(
e^{i\hat{\theta}}\hat{a}_{+1}-e^{-i\hat{\theta}}\hat{a}_{-1}\right)  \right]
+\mathrm{H.c.} \label{quantumXrotating}%
\end{equation}
where $\hat{\theta}$ is half the phase difference operator \cite{Luis} between
opposite angular momentum modes whose exponential form is given\ in terms of
the boson operators by \cite{Yu}%
\begin{equation}
\hat{U}=\exp\left(  i\hat{\theta}\right)  =\left[  \hat{U}_{+1}^{\dagger}%
\hat{U}_{-1}+%
%TCIMACRO{\dsum \limits_{n=0}^{\infty}}%
%BeginExpansion
{\displaystyle\sum\limits_{n=0}^{\infty}}
%EndExpansion
\left\vert 0,n\right\rangle \left\langle n,0\right\vert e^{i\phi\left(
n\right)  }\right]  ^{1/2} \label{ExpOri}%
\end{equation}
being%
\begin{equation}
\hat{U}_{j}=\frac{1}{\sqrt{\hat{a}_{j}^{\dagger}\hat{a}_{j}+1}}\hat{a}_{j}
\label{S-G}%
\end{equation}
the Susskind-Glogower phase operator of the mode\ $j$, the state $\left\vert
m,n\right\rangle =\left\vert m\right\rangle _{+1}\otimes\left\vert
n\right\rangle _{-1}$ a vector of the number state basis for the joined
Hilbert space of both modes ($\left\langle m,n\right\vert =\left\langle
m\right\vert _{+1}\otimes\left\langle n\right\vert _{-1}$) and $\phi\left(
n\right)  $ an arbitrary function defined on the natural numbers. For reasons
that should be clear from the Introduction above, we will call $\hat{\theta}$
the \textit{orientation} operator, which is simply given by%
\begin{equation}
\hat{\theta}=\frac{1}{i}\ln\hat{U}\text{.} \label{Ori}%
\end{equation}

Now the way to follow seems clear; we want to prove that $\hat{X}%
^{\psi_{\mathrm{L}}}$ and $\hat{X}^{\psi_{\mathrm{L}}+\frac{\pi}{2}}$ do not
form a canonical pair, while $\hat{X}^{\psi_{\mathrm{L}}}$ and $\hat{\theta}$ do.

On the other hand, two operators $\hat{F}$ and $\hat{G}$ are canonically
related if they satisfy a conmutation relation of the kind%
\begin{equation}
\left[  \hat{F},\hat{G}\right]  =iC \label{CanComRel}%
\end{equation}
where $C$ is a real non-zero number ($C=2$ for usual orthogonal quadratures
$\hat{X}^{\varphi}$ and $\hat{X}^{\varphi+\frac{\pi}{2}}$). Hence, if we can
prove that%
\begin{equation}
\left[  \hat{X}^{\psi_{\mathrm{L}}},\hat{X}^{\psi_{\mathrm{L}}+\frac{\pi}{2}%
}\right]  =0\text{ and }\left[  \hat{X}^{\psi_{\mathrm{L}}},\hat{\theta
}\right]  =iC \label{DesCom}%
\end{equation}
we will give the proof we are searching for.

However, there is one easier way to prove if two operators are canonically
related or not: instead of using Quantum Field Theory (QFT) one can just prove
whether two observables are canonically related or not via Poisson brackets in
a Classical Field Theory (CFT) context. This idea is the one we develop in the
next part.

\textit{The clasical field theory resort}. The usual approach one uses to move
from CFT to QFT is to change the classical normal variables for each mode of
the field, $\alpha_{j}$ and $\alpha_{j}^{\ast}$, by boson operators $\hat
{a}_{j}$ and $\hat{a}_{j}^{\dagger}$, satisfying conmutation relations%
\begin{equation}
\left[  \hat{a}_{j},\hat{a}_{j}^{\dagger}\right]  =i\left\{  \alpha_{j}%
,\alpha_{j}^{\ast}\right\}  \label{CanbraToCancom}%
\end{equation}
where $\left\{  F,G\right\}  $ denotes the poisson bracket between two
functions $F\left(  \alpha_{j},\alpha_{j}^{\ast}\right)  $ and $G\left(
\alpha_{j},\alpha_{j}^{\ast}\right)  $ defined as%
\begin{equation}
\left\{  F,G\right\}  =\frac{1}{i}%
%TCIMACRO{\dsum \limits_{j}}%
%BeginExpansion
{\displaystyle\sum\limits_{j}}
%EndExpansion
\frac{\partial F}{\partial\alpha_{j}}\frac{\partial G}{\partial\alpha
_{j}^{\ast}}-\frac{\partial F}{\partial\alpha_{j}^{\ast}}\frac{\partial
G}{\partial\alpha_{j}}. \label{BraDef}%
\end{equation}

It can be checked out that by applying this definition onto the fundamental
brackets $\left\{  \alpha_{j},\alpha_{j}^{\ast}\right\}  $ makes
(\ref{CanbraToCancom}) coincide with the canonical conmutation relations
(\ref{CanCom}).

The Poisson bracket of two monomode orthogonal quadratures $X_{j}=\alpha
_{j}+\alpha_{j}^{\ast}$ and $Y_{j}=-i\left(  \alpha_{j}-\alpha_{j}^{\ast
}\right)  $ is then%
\begin{equation}
\left\{  X_{j},Y_{j}\right\}  =2, \label{XYbra}%
\end{equation}
as expected by the conmutation relation (\ref{CanComRel}). In general two
classical functions $F\left(  \alpha_{j},\alpha_{j}^{\ast}\right)  $ and
$G\left(  \alpha_{j},\alpha_{j}^{\ast}\right)  $ are said to form a canonical
pair if their Poisson bracket is of the kind%
\begin{equation}
\left\{  F,G\right\}  =C \label{CanBraRel}%
\end{equation}
where $C$ is again a real number. Hence, we don't need to calculate the
difficult conmutators (\ref{DesCom}) to prove what we want, we can just
compute the analogous Poisson brackets (which only need making some
derivatives) for functions defined in a context of CFT.

In particular, the functions we are interested in, are the classical
counterparts of (\ref{quantumXrotating}) and (\ref{Ori}) which are given by%
\begin{equation}
X^{\psi_{\mathrm{L}}}=\frac{i}{\sqrt{2}}\left[  e^{-i\psi_{L}}\left(
e^{i\theta}\alpha_{+1}-e^{-i\theta}\alpha_{-1}\right)  \right]  +\mathrm{c.c.}
\label{classicalXrotating}%
\end{equation}
with%
\begin{equation}
e^{i\theta}=\frac{\alpha_{+1}^{\ast}\alpha_{-1}}{\sqrt{\alpha_{+1}^{\ast
}\alpha_{+1}}\sqrt{\alpha_{-1}^{\ast}\alpha_{-1}}}, \label{ClassExpOri}%
\end{equation}
and%
\begin{equation}
\theta=\frac{1}{i}\ln\left[  \frac{\alpha_{+1}^{\ast}\alpha_{-1}}{\sqrt
{\alpha_{+1}^{\ast}\alpha_{+1}}\sqrt{\alpha_{-1}^{\ast}\alpha_{-1}}}\right]  .
\label{ClassOri}%
\end{equation}

If we can prove that%
\begin{equation}
\left\{  X^{\psi_{\mathrm{L}}},X^{\psi_{\mathrm{L}}+\frac{\pi}{2}}\right\}
=0\text{ and }\left\{  X^{\psi_{\mathrm{L}}},\theta\right\}  =C \label{DesBra}%
\end{equation}
we will prove what we are trying too.

Using the definition of the Poisson brackets and after some algebra, it is
posible to show that%
\begin{equation}
\left\{  X^{\psi_{\mathrm{L}}},X^{\psi_{\mathrm{L}}+\frac{\pi}{2}}\right\}
=\frac{\left\vert \alpha_{-1}\right\vert -\left\vert \alpha_{+1}\right\vert
}{2\left\vert \alpha_{+1}\right\vert \left\vert \alpha_{-1}\right\vert }
\label{DesBra1}%
\end{equation}
and%
\begin{equation}
\left\{  X^{\psi_{\mathrm{L}}},\theta\right\}  =\frac{i\left(  \left\vert
\alpha_{+1}\right\vert +\left\vert \alpha_{-1}\right\vert \right)  \left(
e^{i\psi_{L}}\left\vert \alpha_{-1}\right\vert -e^{-i\psi_{L}}\left\vert
\alpha_{-1}\right\vert \right)  }{4\sqrt{2}\left\vert \alpha_{+1}\right\vert
^{3/2}\left\vert \alpha_{-1}\right\vert ^{3/2}} \label{DesBra2}%
\end{equation}
with $\left\vert \alpha_{j}\right\vert =\sqrt{\alpha_{j}^{\ast}\alpha_{j}}$.

On the other hand, in the DOPO that was treated in \cite{PRL}, the number of
photons with opposite angular momentum is sensibly equal, i.e., $\left\vert
\alpha_{+1}\right\vert \approx\left\vert \alpha_{-1}\right\vert $; hence, the
dominant term of the previous brackets will be that with $\left\vert
\alpha_{+1}\right\vert =\left\vert \alpha_{-1}\right\vert =\rho$, and thus%
\begin{equation}
\left\{  X^{\psi_{\mathrm{L}}},X^{\psi_{\mathrm{L}}+\frac{\pi}{2}}\right\}
\approx0 \label{Bra1dominant}%
\end{equation}
and%
\begin{equation}
\left\{  X^{\psi_{\mathrm{L}}},\theta\right\}  \approx-\frac{\sin
\psi_{\mathrm{L}}}{\sqrt{2}\rho}\text{.} \label{Bra2Dominant}%
\end{equation}

By comparing this with (\ref{DesBra}) we find that this is a confirmation of
what we expected: the classical field variables $X^{\psi_{\mathrm{L}}}$ and
$X^{\psi_{\mathrm{L}}+\frac{\pi}{2}}$ do not form a canonical pair (moreover,
they commute), while $X^{\psi_{\mathrm{L}}}$ and $\theta$ do. This can be seen
as an indirect proof of the same conclusion for the operators $\hat{X}%
^{\psi_{\mathrm{L}}}$, $\hat{X}^{\psi_{\mathrm{L}}+\frac{\pi}{2}}$ and
$\hat{\theta}$.

\textit{Conclusions}. In these notes we have proven that the apparent
violation of the uncertainty principle that shows the spectrum (\ref{Spectrum}%
) obtained in \cite{PRL} is just that: apparent. Two detected orthogonal
quadratures are not canonically related as the orientation diffusion has been
removed from the experiment by transferring it to the local oscillator field.
Hence, quadratures can be squeezed without interchanging noise between each
other as all the extra noise is carried by the orientation angle $\theta$,
which is the canonically related variable of all the squeezed quadratures. We
have given here a simple proof based on classical field theory calculus.

\end{document}